\documentclass[sigconf]{acmart}
\AtBeginDocument{%
  }

\setcopyright{acmlicensed}

\copyrightyear{2026}
\acmYear{2026}
\setcopyright{cc}
\setcctype{by}
\acmConference[KDD '26]{Proceedings of the 32nd ACM SIGKDD Conference on Knowledge Discovery and Data Mining V.2}{August 09--13, 2026}{Jeju Island, Republic of Korea}
\acmBooktitle{Proceedings of the 32nd ACM SIGKDD Conference on Knowledge Discovery and Data Mining V.2 (KDD '26), August 09--13, 2026, Jeju Island, Republic of Korea}
\acmDOI{10.1145/3770855.3818447}
\acmISBN{979-8-4007-2259-2/2026/08}

\usepackage{enumitem}
\usepackage{bm}
\usepackage{booktabs}
\usepackage{multirow}
\usepackage{ulem}
\usepackage[table]{xcolor}

\begin{document}

\title{MixFormer: Co-Scaling Up Dense and Sequence in Industrial Recommenders}

\author{Xu Huang}
\authornote{Equal contribution.}
\email{xu.hwang@outlook.com}
\orcid{0000-0003-4354-334X}
\affiliation{%
  \institution{ByteDance Inc.}
  \city{Shanghai}
  \country{China}
}

\author{Hao Zhang}
\authornotemark[1]
\email{zhanghao.229@bytedance.com}
\affiliation{%
  \institution{ByteDance Inc.}
  \city{Shanghai}
  \country{China}
}

\author{Zhifang Fan}
\authornotemark[1]
\email{fanzhifangfzf@gmail.com}
\affiliation{%
  \institution{ByteDance Inc.}
  \city{Shanghai}
  \country{China}
}

\author{Yunwen Huang}
\authornotemark[1]
\email{huangyunwen.eleanor@bytedance.com}
\affiliation{%
  \institution{ByteDance Inc.}
  \city{Beijing}
  \country{China}
}

\author{Zhuoxing Wei}
\email{weizhuoxing@bytedance.com}
\affiliation{%
  \institution{ByteDance Inc.}
  \city{Beijing}
  \country{China}
}

\author{Zheng Chai}
\email{chaizheng.cz@bytedance.com}
\affiliation{%
  \institution{ByteDance Inc.}
  \city{Hangzhou}
  \country{China}
}

\author{Jinan Ni}
\email{nijinan@bytedance.com}
\affiliation{%
  \institution{ByteDance Inc.}
  \city{Shanghai}
  \country{China}
}

\author{Yuchao Zheng}
\email{zhengyuchao.yc@bytedance.com}
\affiliation{%
  \institution{ByteDance Inc.}
  \city{Hangzhou}
  \country{China}
}


\author{Qiwei Chen}
\authornotemark[2]
\email{chenqiwei05@gmail.com}
\affiliation{%
  \institution{ByteDance Inc.}
  \city{Shanghai}
  \country{China}
}

\renewcommand{\shortauthors}{Xu Huang et al.}


\begin{abstract}
As industrial recommender systems enter a scaling-driven regime, Transformer architectures have become increasingly attractive for scaling models towards larger capacity and longer sequence. However, existing Transformer-based recommendation models remain structurally fragmented, where sequence modeling and feature interaction are implemented as separate modules with independent parameterization. Such designs introduce a fundamental co-scaling challenge, as model capacity must be suboptimally allocated between dense feature interaction and sequence modeling under a limited computational budget.
In this work, we propose \textbf{MixFormer}, a unified Transformer-style architecture tailored for recommender systems, which jointly models sequential behaviors and feature interactions within a single backbone. Through a unified parameterization, MixFormer enables effective co-scaling across both dense capacity and sequence length, mitigating the trade-off observed in decoupled designs. Moreover, the integrated architecture facilitates deep interaction between sequential and non-sequential representations, allowing high-order feature semantics to directly inform sequence aggregation and enhancing overall expressiveness. To ensure industrial practicality, we further introduce a user-item decoupling strategy for efficiency optimizations that significantly reduce redundant computation and inference latency.
Extensive experiments on large-scale industrial datasets demonstrate that MixFormer consistently exhibits superior accuracy and efficiency. Furthermore, large-scale online A/B tests on two production recommender systems, Douyin and Douyin Lite, show consistent improvements in user engagement metrics, including active days and in-app usage duration.
\end{abstract}




\maketitle

\section{Introduction}

Transformer architectures~\cite{transformer,gpt,bert} have become a foundational modeling paradigm in modern industrial recommender systems. Owing to their strong representation capacity and high degree of parallelism, Transformers are particularly effective for large-scale user behavior modeling and complex feature interaction learning. Meanwhile, recommender systems have entered a scaling-driven regime, where performance gains are increasingly achieved by expanding data volume and model capacity, rather than relying on handcrafted features or task-specific heuristics~\cite{rankmixer,longer,onerec,wukong,dhen,hstu}. Under this regime, model architecture is no longer a secondary design choice but a primary determinant of how efficiently additional computational resources can be translated into measurable performance improvements.

Existing applications of Transformers in recommender systems broadly follow two complementary yet largely independent directions. One line of work focuses on sequential modeling, leveraging Transformers to capture long-range temporal dependencies in user behavior sequences~\cite{longer,din,bst,ETA,sim,ubr}. Another line emphasizes high-order feature interaction, adopting Transformer-style architectures to model complex relationships among heterogeneous dense features~\cite{wukong,rankmixer,dhen}. While both directions have independently demonstrated strong empirical gains and favorable scaling trends, they implicitly assume that sequence modeling and feature interaction can be optimized in isolation. This assumption becomes increasingly fragile as recommender systems scale to longer user histories, richer feature spaces, and larger model capacities.

From a system-level perspective, jointly scaling sequence modeling and dense feature interaction under limited computational budgets exposes a fundamental architectural tension. Sequential Transformers incur computational costs that grow rapidly with sequence length, while dense Transformers primarily scale with feature dimensionality and model width. When these two components are parameterized and optimized separately, they compete for a shared computation and parameter budget, leading to conflicting scaling incentives. Allocating more capacity to the sequence component improves temporal modeling but disproportionately increases computational cost, effectively suppressing the scaling potential of dense feature interaction. Conversely, prioritizing dense scaling limits the model's ability to exploit long-range behavioral signals. As a result, achieving globally optimal co-scaling of sequence and dense components becomes structurally misaligned with architectures that enforce rigid parameter separation.

Most existing approaches address this challenge by combining a sequence Transformer and a dense Transformer through hierarchical stacking or parallel concatenation. In hierarchical designs, the output of the sequence Transformer is treated as an input feature for the dense Transformer~\cite{guan2025make,rankmixer,longer}, whereas parallel designs concatenate the two components~\cite{onetrans}. Despite their simplicity, these paradigms preserve a strict separation between sequence and non-sequence parameters, resulting in limited cross-component interaction and fragmented optimization. More importantly, such separation fundamentally constrains co-scaling behavior: the rapidly growing computational footprint of the sequence module dominates resource allocation decisions, preventing the dense component from scaling proportionally and leading to suboptimal global performance under realistic deployment constraints.

In this work, we argue that effective co-scaling in large-scale recommender systems requires a fundamentally different architectural principle: sequence modeling and feature interaction must be unified within a single parameter space and jointly optimized throughout the entire network. To this end, we propose \textbf{MixFormer}, a fully unified Transformer-style Large Recommender Model that employs a shared set of parameters to simultaneously model sequence and dense feature interactions. By eliminating rigid parameter boundaries, MixFormer enables sequential and non-sequential co-modeling to mutually enhance each other, allowing high-order feature semantics to directly participate in sequential aggregation while preserving fine-grained behavioral signals. This unified parameterization fundamentally resolves the parameter allocation dilemma and establishes a coherent foundation for co-scaling.

Furthermore, to make unified scaling practically viable in industrial environments, we introduce a user-item decoupling strategy that leverage request-level batching~\cite{guan2025make} technique to reuse computations, significantly improving computational efficiency. This mechanism is a critical enabler for scaling unified Transformer architectures under constrained resource budgets, effectively raising the ceiling of model capacity and sequence length that can be deployed in production systems.

In summary, our contributions are threefold:
\begin{itemize}[leftmargin=*]
\item We propose MixFormer, a fully unified Transformer architecture for recommender systems that jointly models sequence dynamics and dense feature interactions within a single parameter space, addressing the structural limitations of existing hybrid designs and enabling effective global co-scaling.
\item We introduce a user-item decoupling strategy for request-level computation sharing and reduction that makes unified Transformer scaling feasible in large-scale industrial settings.
\item Extensive experiments on large-scale industrial datasets demonstrate that MixFormer achieves superior accuracy and efficiency and exhibits more favorable co-scaling behavior with increasing model capacity and sequence length.
\end{itemize}

\section{Related Work}

\subsection{Sequence Modeling}
Modeling user behavioral sequences is a fundamental problem in recommender systems, as it enables the capture of users' dynamic and evolving interests. Early and representative approaches focus on modeling recent user actions using target-attention-based architectures, including DIN~\cite{din}, DIEN~\cite{DIEN}, and BST~\cite{bst}. These methods typically emphasize short-term user interests by attending to a limited window of recent behaviors. However, such short sequences are often insufficient to fully characterize long-term user preferences.
To address this limitation, SIM~\cite{sim} introduces a retrieval-based paradigm for extremely long sequence modeling. Specifically, SIM employs a two-stage framework consisting of a General Search Unit (GSU) to retrieve relevant historical behaviors and an Extract Search Unit (ESU) to model the retrieved subsequence. Building upon this paradigm, subsequent works further optimize SIM by enabling end-to-end training~\cite{ETA}, enhancing the retrieval strategy with BM25-based algorithms~\cite{ubr}, and improving the GSU module with consistency-preserving mechanisms~\cite{TWIN}, among others~\cite{TWINv2,MIRRN,SDIM}.
With the rapid advancement of GPU-based hardware and large-scale training infrastructure, recent studies have revisited end-to-end long sequence modeling. Longer~\cite{longer} proposes a hierarchical attention architecture to reduce the quadratic complexity of self-attention over long sequences. Meanwhile, GR~\cite{hstu} and MTGR~\cite{MTGR} reformulate click-through rate (CTR) prediction as a generative task by adopting Transformer-decoder-style architectures.
However, existing sequence modeling methods largely concentrate on sequential signals alone, risking insufficient query expressiveness and the loss of fine-grained behavioral information.

\subsection{Feature Interaction}
Feature interaction has long been a core research topic in recommender systems, aiming to construct high-order representations by combining heterogeneous input features. Factorization Machines~\cite{FM} are among the earliest approaches to explicitly model second-order feature interactions, followed by a wide range of extensions that explore higher-order interactions and implicit cross features using neural networks~\cite{DeepFM,xDeepFM,DCN,DCNv2}.
However, recent studies have shown that many existing neural interaction models struggle to scale effectively to industrial-scale recommendation scenarios~\cite{wukong}. To address this challenge, WuKong~\cite{wukong} is proposed as a large-scale feature interaction backbone, demonstrating strong scalability and expressive power. Building on this line of work, RankMixer~\cite{rankmixer} further introduces a high-efficiency, Transformer-style architecture tailored for large-scale industrial ranking systems.
Despite their success, these feature interaction methods typically treat sequential features as compressed or static representations. Consequently, the high-order features produced by interaction modules are decoupled from the sequence modeling process, limiting the expressiveness of sequential representations.

Industrial recommender systems commonly adopt a hierarchical stacking or parallel combination paradigm to combine sequential and non-sequential modules. Recently, OneTrans~\cite{onetrans} attempts to unify the two by modeling sequential and non-sequential features as a heterogeneous token sequence within a Transformer backbone using designed attention masks and independent parameters. However, the quadratic complexity introduces severe computational overhead, and the departed parameters lead to the challenge of co-scaling of dense capacity and sequence length.

\section{Methodology}

\subsection{Overall}
\begin{figure*}[t]
    \centering
    \includegraphics[width=\linewidth]{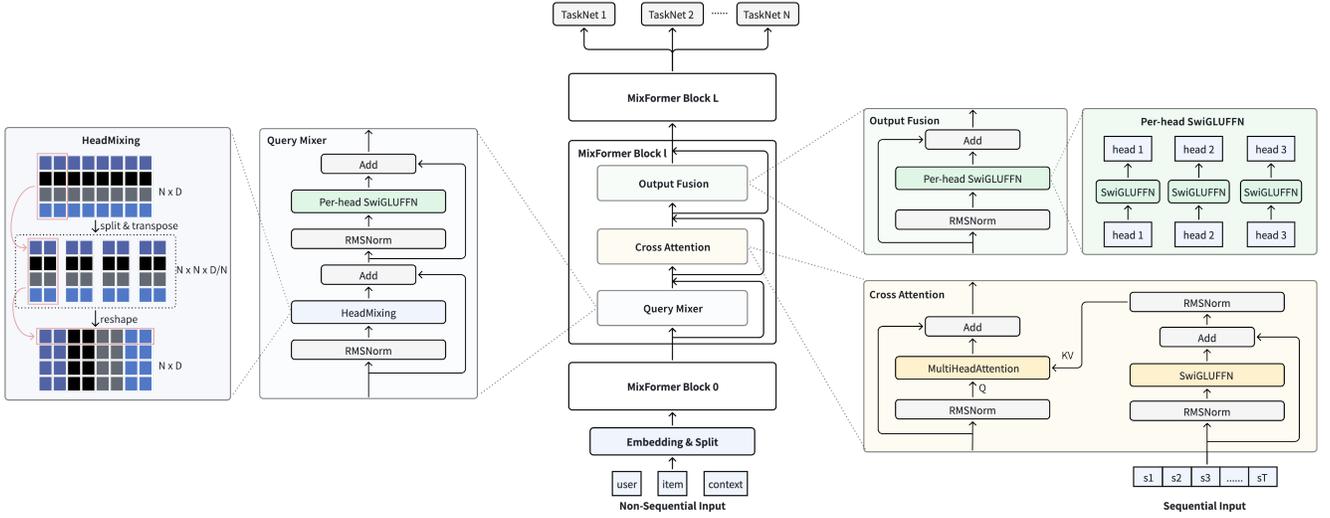}
    \caption{Overview of the MixFormer framework. The architecture consists of a feature embedding and split layer followed by $L$ stacked MixFormer blocks and task-specific heads. Each MixFormer block integrates three core components: a Query Mixer, Cross-Attention, and Output Fusion.}
    \Description{Overview of the MixFormer framework. The architecture consists of a feature embedding and split layer followed by $L$ stacked MixFormer blocks and task-specific heads. Each MixFormer block integrates three core components: a Query Mixer, Cross-Attention, and Output Fusion.}
    \label{fig:framework}
\end{figure*}

MixFormer is an efficient variant of decoder-only Transformer tailored for multi-task recommender systems, consisting of an input layer and $L$ MixFormer blocks, followed by several task networks. The input layer is responsible for feature embedding and split, which first projects user, item, and context features into an embedding vector and then splits the vector into $N$ heads. 
The overall architecture is presented in Figure~\ref{fig:framework}.


\subsection{Feature Embedding and Splitting}\label{sec:feat_emb_token}
Input features can be separated into two categories: \textit{sequential features} and \textit{non-sequential features}. 

Sequential features represent a user's historical interaction sequence, consisting of temporally ordered user actions on items. Each action is characterized by an item identifier, an action type, a timestamp, and some available side information.
For each action at time step $t$, we embed its constituent features using dedicated embedding layers and concatenate the resulting vectors to form an action representation $\bm{s}_t$. 
Accordingly, a user behavior sequence of length $T$ is represented as
\[
\bm{S} = [\bm{s}_1, \bm{s}_2, \cdots, \bm{s}_T],
\]
which serves as the sequential input to the model.

The non-sequential features include user features, item features, and contextual features.
Let $\mathcal{F}_{\text{ns}} = \{f_1, f_2, \dots, f_M\}$ denote the set of non-sequential features. Each feature $f_i$ is first mapped into a dense embedding vector $\bm{e}_i \in \mathbb{R}^{d_i}$ through a feature-specific embedding table. These embeddings are then concatenated into a single composite representation:
\begin{equation}
\mathbf{e}_{\text{ns}} = [\bm{e}_1; \bm{e}_2; \dots; \bm{e}_M] \in \mathbb{R}^{D_{\text{ns}}},
\end{equation}
where $D_{\text{ns}} = \sum_{i=1}^{M} d_i$ denotes the total embedding dimension of all non-sequential features. The non-sequential features then serve as the query input for the entire backbone.

In a standard Transformer decoder, the attention modules (self attention or cross attention) are extended to a multi-head version, where the query are split in multiple subspaces. To enable unified modeling within Transformer-style architectures, we further split $\mathbf{e}_{\text{ns}}$ into a set of fixed-dimensional feature heads. Specifically, $\mathbf{e}_{\text{ns}}$ is evenly partitioned into $N$ contiguous subvectors, each of dimension $d = D_{\text{ns}} / N$. Each subvector is then projected into a $D$-dimensional vector:
\begin{equation}
\bm{x}_j = \bm{W}_j\cdot \bm{e}_{\text{ns}}[d\cdot(j-1) : d\cdot j],
\quad j = 1, \dots, N,
\end{equation}
where $\bm{W}_j\in \mathbb{R}^{D\times d}$ denotes a learnable linear matrix, and $\bm{x}_j \in \mathbb{R}^{D}$ represents the $j$-th non-sequential feature head. The resulting vector $\bm{X}=[{\bm{x}_1, \dots, \bm{x}_N}]\in\mathbb{R}^{ND}$ serves as the non-sequential input head to MixFormer blocks. 

Compared to collapsing all features into a single head, partitioning the embedding space into multiple heads preserves representational diversity, enabling the model to capture heterogeneous feature semantics without introducing excessive structural complexity. Besides, three kernel modules (Query Mixer, Cross Attention and Output Fusion) in the MixFormer block introduced in following sections are all designed in a multi-head manner. This design provides a flexible and computation-efficient interface for downstream blocks, allowing high-order feature interactions to be seamlessly integrated with sequence modeling in later stages.


\subsection{MixFormer Block}
MixFormer adopts a Transformer-style architecture that is structurally aligned with the Transformer decoder. 
Each MixFormer block consists of three core modules: \emph{Query Mixer}, \emph{Cross Attention}, and \emph{Output Fusion}, which conceptually correspond to the Self-Attention, Cross-Attention, and Feed-Forward Network (FFN) components in a standard Transformer decoder block. 
This design preserves the expressive capacity of Transformer-based architectures while enabling customized operations tailored to large-scale industrial recommendation systems.

\subsubsection{Query Mixer}
Self-attention has demonstrated remarkable effectiveness in large language models, where all tokens are embedded in a unified semantic space and token-to-token similarity can be reliably modeled through inner-product operations. 
However, inspired by RankMixer~\cite{rankmixer}, the query are derived from highly heterogeneous feature fields in recommendation scenarios, including user attributes, item attributes, and contextual signals. 
These features originate from distinct semantic spaces and often correspond to extremely large and sparse ID domains. Under such heterogeneity, computing attention weights via inner-product similarity becomes inherently unreliable, as meaningful alignment between different feature spaces is difficult to establish. 
As a result, self-attention not only fails to consistently improve modeling effectiveness, but also introduces substantial computational overhead.

Therefore, we replace self-attention with a lightweight and hardware-friendly \emph{Query Mixer} module inspired by prior works~\cite{rankmixer, mlpmixer}.
Similar to multi-head self-attention module, the Query Mixer enables cross-head information exchange without relying on similarity-based attention, making it better suited for heterogeneous feature modeling under strict efficiency constraints. 

Formally, given the input query $\bm{X}=[\bm{x}_1, \bm{x}_2, \cdots, \bm{x}_N]\in \mathbb{R}^{N\times D}, \bm{x}_i \in \mathbb{R}^d$.
Then the Query Mixer is defined as:
\begin{align}
    \bm{P} &= [\bm{p}_1, \cdots,\bm{p}_N]= \operatorname{HeadMixing}(\operatorname{Norm}(\bm{X})) + \bm{X}, 
    \label{TokenMixing}
    \\
    \bm{q}_i &= \operatorname{SwiGLUFFN}_i(\operatorname{Norm}(\bm{p}_i)) + \bm{p}_i.
\end{align}
Here, $\operatorname{HeadMixing}(\cdot)$ reshapes $X$' into $\mathbb{R}^{N \times N \times \frac{D}{N} }$, transposes the first and second dimensions, and flattens the result back to $\mathbb{R}^{N \times D}$, which is illustrated in Figure~\ref{fig:framework}. 
This operation enables efficient cross-head information exchange in a parameter-free manner.
Notably, the FFN is instantiated independently for each head to explicitly account for feature heterogeneity, which is named as per-head FFN.
This design has been shown to provide strong expressive capacity while maintaining favorable efficiency in the prior work~\cite{rankmixer}.

\subsubsection{Cross Attention}
While similarity-based self-attention is suboptimal for modeling heterogeneous non-sequential features, it remains an effective mechanism for aligning structured query representations with sequential behavioral signals. 
In MixFormer, the Cross Attention module is designed to aggregate user sequences conditioned on high-order feature representations produced by the Query Mixer.
Specifically, the $N$ output heads from the Query Mixer are directly treated as $N$ heads for cross attention, where each head serves as a semantically specialized sub-query that focuses on a distinct subspace of non-sequential features. 
This design enables different aspects of user preferences to attend to behavioral sequences in a disentangled yet coordinated manner, avoiding the need for additional projection matrices for query splitting and reducing the risk of head collapse.

We first transform each action in the behavioral sequence into a latent vector $\bm{h}_t$ to align with the query input using a per-layer SwiGLU-activated FFN, and then project it into key and value:
\begin{align}
\bm{h}_t &= \operatorname{SwiGLUFFN}^{(l)}(\operatorname{Norm}(\bm{s}_t))+\bm{s}_t \in \mathbb{R}^{ND}, \\
\bm{h}_{t}^{i} &= \bm{h}_t[iD:(i+1) D] \in \mathbb{R}^D,\\
\bm{k}_{t}^{i} &= \bm{W}_k^i\bm{h}_{t}^{i}; \quad \bm{v}_t = \bm{W}_v^i\bm{h}_t^i
\end{align}
where $\operatorname{SwiGLU}^{(l)}$ denotes SwiGLUFFN in the $l$-th MixFormer block, respectively. $\bm{W}_k^i,\bm{W}_v^i\in\mathbb{R}^{D\times D}$ denote the projection matrix for keys and values for the $i$-th head.
Unlike standard cross-attention mechanisms in the Transformer decoder that rely on shared hiddens, these per-layer SwiGLUFFN are independently parameterized at each layer, allowing the model to progressively refine sequence representations across depth.

The output for the $i$-th query head is computed as:
\begin{equation}
\bm{z}_i = \sum_{t=1}^{T} 
\operatorname{softmax}\!\left(
\frac{\bm{q}_i^\top \bm{k}_t^i}{\sqrt{D}}
\right)
\bm{v}_t^i +\bm{q}_i,
\quad i = 1, \dots, N.
\end{equation}
The resulting representations $\{\bm{z}_1, \dots, \bm{z}_N\}$ capture feature-conditioned summaries of the user sequence and serve as the aggregated sequential outputs for subsequent fusion.

\subsubsection{Output Fusion}
After obtaining high-order non-sequential representations from the Query Mixer and feature-conditioned sequential aggregations from the Cross Attention module, the Output Fusion layer performs deep integration of these signals to produce the final representations.

Specifically, each cross-attention output $\bm{z}_i$ contains both non-sequential information and the sequential information aligned with the corresponding high-order query head. 
However, due to the heterogeneous nature of query head and the residual connections introduced in previous modules, a shared feed-forward transformation is insufficient to fully capture head-specific interactions. 
Therefore, we adopt a per-head SwiGLU-activated Feed-Forward Network to further refine each representation independently.

Formally, the Output Fusion is defined as:
\begin{equation}
    \bm{o}_i = \operatorname{SwiGLUFFN}_i(\operatorname{Norm}(\bm{z}_i)) + \bm{z}_i,
\end{equation}
where $\operatorname{SwiGLUFFN}_i(\cdot)$ denotes a head-specific SwiGLUFFN

This design enables MixFormer to perform deep, non-linear fusion of sequential and non-sequential signals while explicitly accounting for feature heterogeneity. 
By applying independent transformations to each head, the model preserves head-level specialization and avoids representational interference across heterogeneous feature subspaces, leading to more expressive and stable representations. 
The output then serves as the input of the next MixFormer block, achieving a progressive unified feature interaction and sequence modeling.
Notably, the per-head FFNs in Output Fusion and Query Mixer serve both sequential and non-sequential features, forming a unified parameterization paradigm and addressing the challenge of allocation of parameters in the combination of two independent modules.

\subsection{User-Item Decoupling}

\begin{figure}[t]
    \centering
    \includegraphics[width=\linewidth]{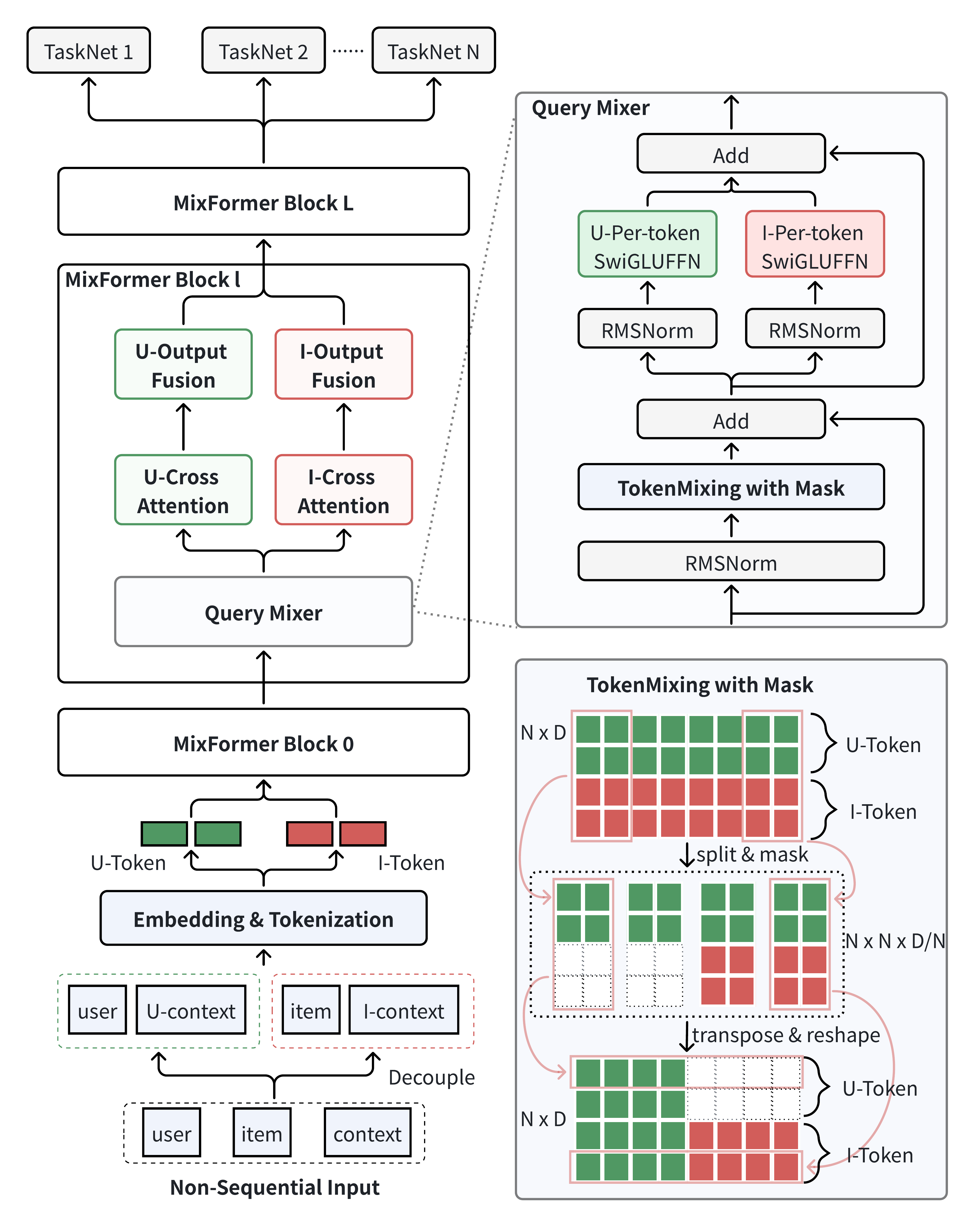}
    \caption{Overview of the User-Item Decoupled MixFormer with a request-level reduction technique. The green modules denote user-side computations that can be request-level shared, while the red modules indicate item-side computations that are performed independently for each candidate item and therefore cannot be shared.}
    \label{fig:framework_ug}
    \Description{Overview of the User-Item Decoupled MixFormer with a request-level reduction technique. The green modules denote user-side computations that can be request-level shared, while the red modules indicate item-side computations that are performed independently for each candidate item and therefore cannot be shared.}
\end{figure}

Request Level Batching (RLB) \cite{guan2025make,roo} emerges as an efficient paradigm for boosting the training and inference efficiency in recommendation, which shares the user-side computation across multiple targets within a single request to achieve substantial reduction in computational cost. 
However, the mixed user-item computations in the original unified MixFormer limit RLB's application. To address this, we propose a User-Item decoupled 
MixFormer variant (\textbf{UI-MixFormer}) as shown in Figure~\ref{fig:framework_ug}.

\subsubsection{Feature Decoupling.}
We partition non-sequential features into disjoint user-side and item-side subsets, projecting them into $N_U$ and $N_G$ heads, respectively, serving as the input to Mixformer. To preserve model capacity, the total head number remains unchanged, and $N_U$ and $N_G$ are calculated according to embedding dimension: $N_G = \lfloor\frac{D_{ns}^G N}{D_{ns}}\rfloor$, $N_U = N - N_G$ (where $D_{ns}^G$ is the embedding dimension of item-side non-sequential features). Practically, $N_U:N_G$ is set to 1:1.

\subsubsection{Query Mixer with Mask.} 
Due to the HeadMixing operation applied across distinct heads, its output no longer retains a pure user-side representation, which renders the acceleration via RLB infeasible.
Inspired by causal masks in Self-Attention, we design a mask for unidirectional user-to-item fusion in Mixformer. As illustrated in the figure \ref{fig:framework_ug}, 
To ensure that the user-side heads can be reused within a single request, we remove the item-side signals of user side heads with a 
mask matrix $\mathcal{M} \in \mathbb{R}^{N  \times D }$ which is defined as:
\begin{equation}
\mathcal{M}[i,j] = 
\begin{cases} 
0,& i < N_U \;and \; j \ge N_U * \frac{D}N\\
1,& otherwise.
\end{cases}
\end{equation}
Finally, element-wise multiplication of the mask matrix 
$\mathcal{M}$
 with the output of the original HeadMixing operation yields a user-item decoupled HeadMixing module, thus realizing a query mixer with explicit user-item decoupling:
\begin{equation}
\operatorname{HeadMixing}_{\text{decouple}}(\cdot) = \mathcal{M} \odot \operatorname{HeadMixing}(\cdot)
\end{equation}
In contrast to dual-tower architectures, the proposed method retains the information propagation from the user side to the item side, thereby ensuring effective modeling of user-item feature interactions. 
It is worth noting that the sequential features of user historical behaviors are also amenable to request-level sharing. As a result, the computational overhead incurred by the cross-attention computation between user-side heads and sequences can be drastically cut down. This observation further highlights the merits of the unified modeling paradigm that that integrates sequence modeling and feature interaction into a unified backbone.

\section{Experiments}

\begin{table*}[t]
\caption{Performance over accuracy and efficiency compared with state-of-the-art baselines($L=512$). The \textbf{bold} and \uline{underlined} numbers represent the best and the second best results, respectively. The \#Params only counts for the dense parameters, including the small modules in input projection and the task networks. The FLOPs are calculated without enabling the request-level batching for fair comparison, except the UI-MixFormer. "-" denotes the results are missing due to platform issues.}
\vspace{-0.3cm}
\label{tab:overall}
\centering
\setlength{\tabcolsep}{8pt}
\begin{tabular}{@{}llrrrrrr@{}}
\toprule
\multirow{2}{*}{\textbf{Type}}  & \multirow{2}{*}{\textbf{Model}} & \multicolumn{2}{c}{\textbf{Finish}} & \multicolumn{2}{c}{\textbf{Skip}} & \multicolumn{2}{c}{\textbf{Effiency}} \\
\cmidrule(l){3-4} \cmidrule(l){5-6} \cmidrule(l){7-8}  
& & \textbf{AUC} $\uparrow$ & \textbf{UAUC} $\uparrow$ & \textbf{AUC} $\uparrow$ & \textbf{UAUC} $\uparrow$ & \textbf{\#Params}(M) & \textbf{GFLOPs/Batch} \\
\midrule
\multirow{8}{*}{\textbf{Stacked}}
&\textbf{TA}$\to$\textbf{DLRM}       & 0.8554 & 0.8270 & 0.8124 & 0.7294 & 9 & 52 \\
&\textbf{TA}$\to$\textbf{DCNv2}     & +0.13\% & +0.13\% & +0.15\% & +0.26\% & 22 & 170 \\
&\textbf{TA}$\to$\textbf{DHEN}      & +0.18\% & +0.26\% & +0.36\% & +0.52\% & 22 & 158 \\ 
&\textbf{TA}$\to$\textbf{Wukong}    & +0.29\% & +0.29\% & +0.49\% & +0.65\% & 122 & 442 \\ 
&\textbf{STCA}$\to$\textbf{DCNv2}  & +0.89\% & +0.91\% & +1.05\% & +1.65\% & 145 & 4,560 \\  
&\textbf{TA}$\to$\textbf{RankMixer}  & +0.95\% & +1.22\% & +1.25\% & +1.82\% & 1,118 & 2,180 \\
&\textbf{STCA}$\to$\textbf{RankMixer}  & \uline{+1.12}\% & \uline{+1.40}\% & \uline{+1.43}\% & \uline{+2.14}\% & 1,255 &  6,736\\ 
\midrule
\multirow{3}{*}{\textbf{Parallel}}
&\textbf{OneTrans}    & +1.05\% & +1.31\% & +1.30\% & +1.95\% & 316 & 23,371  \\
&\textbf{STCA}$\oplus$\textbf{RankMixer}    & +1.11\% & +1.38\% & +1.42\% & +2.11\% & 1,255 &  6,736 \\
\midrule
\rowcolor{cyan!15}
&\textbf{MixFormer-small} & +1.01\% & -- & +1.18\% & -- & 282 & 733 \\ 
\rowcolor{cyan!15}
&\textbf{MixFormer-medium}   & \textbf{+1.28}\% & \textbf{+1.60}\% & \textbf{+1.60}\% & \textbf{+2.46}\% & 1,226 & 3,503 \\
\rowcolor{cyan!15}
\multirow{-3}{*}{\textbf{Unified}}
&\textbf{UI-MixFormer-medium}   & \textbf{+1.28}\% & \textbf{+1.60}\% & \textbf{+1.60}\% & \textbf{+2.46}\% & 1,226 & 2,242 \\ 
\bottomrule
\end{tabular}
\end{table*}

\subsection{Experimental Setting}
\subsubsection{Dataset}
We conduct experiments on a large-scale offline dataset collected from the Douyin recommendation system. The dataset spans two consecutive weeks and contains trillions of user-item interaction records. Each instance is associated with over 300 features, which can be broadly categorized into \emph{non-sequential} and \emph{sequential} features. Non-sequential features include categorical, numerical, and cross features derived from user profiles, item attributes, and contextual information. Sequential features correspond to users' histories, where each action is represented by an item identifier, action type, timestamp, and side attributes.

\subsubsection{Evaluation Metrics}
We evaluate model performance from both accuracy and efficiency perspectives. For accuracy evaluation, we formulate the task as click-through rate (CTR) prediction, which is a binary classification problem. We adopt the area under the ROC curve (AUC) and user-level AUC (UAUC) as the primary evaluation metrics, as they are widely used in industrial recommendation systems to assess ranking quality and user-level consistency. For efficiency evaluation, we report the number of model parameters and floating-point operations (FLOPs), which serve as proxies for computational complexity and are indicative of a model's scalability and deployment cost in real-world systems.

\subsubsection{Baselines}
We compare MixFormer against two categories of state-of-the-art baselines.  
\textbf{(1) Stacked methods}, which first aggregate sequential features into a compact representation and then feed it into a dedicated feature interaction module. For sequence modeling, we adopt several variants, including simple target attention(TA) and the state-of-the-art Stacked Target Cross Attention (STCA) architecture~\cite{guan2025make}. For feature interaction, we consider several representative methods, including DLRM (the vanilla MLP), DCNv2~\cite{DCNv2}, Wukong~\cite{wukong}, DHEN~\cite{dhen}, and RankMixer~\cite{rankmixer}. "A$\to$B" denotes that the output of A is treated as the input of B.
\textbf{(2) Parallel methods}, which concatenate the non-sequential and sequential modules in parallel, where the parameters of two modules are independent. We select RankMixer$\oplus$STCA and OneTrans~\cite{onetrans} as representative baselines in this category. "A$\oplus$B" denotes that the outputs of A and B are concatenated and then fed into the task networks.
As for MixFormer, we conduct experiments with two sizes, denoted as \textbf{MixFormer-small} and \textbf{MixFormer-medium}.

\subsubsection{Implementation Details}
All experiments were conducted on hundreds of GPUs in a hybrid distributed training framework that the sparse part is updated asynchronously, while the dense part is updated synchronously. The optimizer hyperparameters were kept consistent across all
models. For the dense part, we used the RMSProp optimizer with a learning rate of 0.01, while the sparse part used the Adagrad optimizer. The batch size is set as 1,500 in all experiments. The hyperparameters are set as $N=16,L=4,D=386$ and $N=16,L=4,D=768$ in MixFormer-small and MixFormer-medium, respectively. 

\begin{table*}[t]
\caption{Online A/B test results for feed recommendation scenarios on the Douyin and Douyin Lite apps. The improvements are calculated by comparing with the most competitive STCA$\to$RankMixer baseline deployed online. All reported improvements are statistically significant and have not yet converged; the results shown here continue to improve.}
\vspace{-0.3cm}
\resizebox{\textwidth}{!}{
\renewcommand\arraystretch{1.08}
\begin{tabular}{lccccc ccccc}
\toprule
 & \multicolumn{5}{c}{\textbf{Douyin app}} & \multicolumn{5}{c}{\textbf{Douyin Lite}} \\
\cmidrule(lr){2-6} \cmidrule(lr){7-11}
 & \textbf{Active Day}$\uparrow$ 
 & \textbf{Duration}$\uparrow$ 
 & \textbf{Like}$\uparrow$ 
 & \textbf{Finish}$\uparrow$ 
 & \textbf{Comment}$\uparrow$ 
 & \textbf{Active Day}$\uparrow$ 
 & \textbf{Duration}$\uparrow$ 
 & \textbf{Like}$\uparrow$ 
 & \textbf{Finish}$\uparrow$ 
 & \textbf{Comment}$\uparrow$ \\
\midrule
\textbf{Overall}       & +0.0415\%           & +0.2799\%         & 
+0.1766\%     & +0.3897\%       & +0.7035\%        & +0.0252\%           & +0.4105\%          & 
+0.2125\%     & +0.2924\%        & +1.9097\%        \\ \hline 
\textbf{Low-active}    & +0.2263\%            & +0.2468\%         & 
+0.0771\%     & +0.4123\%       & +1.2483\%         & +0.2543\%           & +0.6044\%         & +3.0565\%      & +0.6157\%       & +2.6452\%        \\
\textbf{Middle-active} & +0.0998\%           & +0.2719\%         & +0.2445\%     & +0.2796\%       & +0.6718\%         & +0.1218\%           & +0.4184\%         & +0.2329\%      & +0.2951\%         & +1.3286\%        \\
\textbf{High-active}   & +0.0203\%           & +0.2938\%         & +0.381\%      & +0.3335\%       & +0.8356\%          & +0.0237\%           & +0.4042\%         & +0.4871\%      & +0.2097\%        & +2.117\% 
\\ 
\bottomrule
\end{tabular}
}
\label{tab:online_ab}
\end{table*}

\subsection{Comparison with Baselines}
To demonstrate the performance in terms of accuracy and efficiency of the proposed unified model, we compare MixFormer with the state-of-the-art methods, including hierarchical stacked methods and two representative methods. The results are shown in Table~\ref{tab:overall}, from which we have following findings:

State-of-the-art approaches for sequence modeling and feature interaction, such as STCA and RankMixer, consistently outperform alternative architectures including TA, DCN, and Wukong by a clear margin. This result empirically verifies the effectiveness of Transformer-style architectures in capturing complex dependencies and modeling large-scale user behavior patterns, highlighting their advantages in industrial recommendation systems.

Under comparable sub-structure configurations, the Stacked and Parallel design paradigms exhibit only marginal performance differences. This observation indicates that when the two components are parameterized separately, the interaction between sequence modeling and feature interaction remains relatively shallow. Consequently, modifying the connection pattern alone is insufficient to produce substantial performance improvements, suggesting that deeper integration between the two modules is necessary to fully exploit their complementary strengths.

Benefiting from a unified architecture, the proposed MixFormer consistently outperforms all baseline methods. The improvements mainly arise from parameter sharing between the feature interaction and sequence modeling components, which facilitates enhanced cross-module representation learning. Notably, under a comparable parameter budget to the strongest baseline STCA + RankMixer, MixFormer achieves significant AUC improvements across both tasks. 
Moreover, the efficiency-optimized variant with User–Item Decoupling substantially reduces computational overhead, achieving $\sim$36\% FLOPs reduction while maintaining nearly identical performance.

\subsection{Ablation Study}
\begin{figure}[t]
    \centering
    \includegraphics[width=0.9\linewidth]{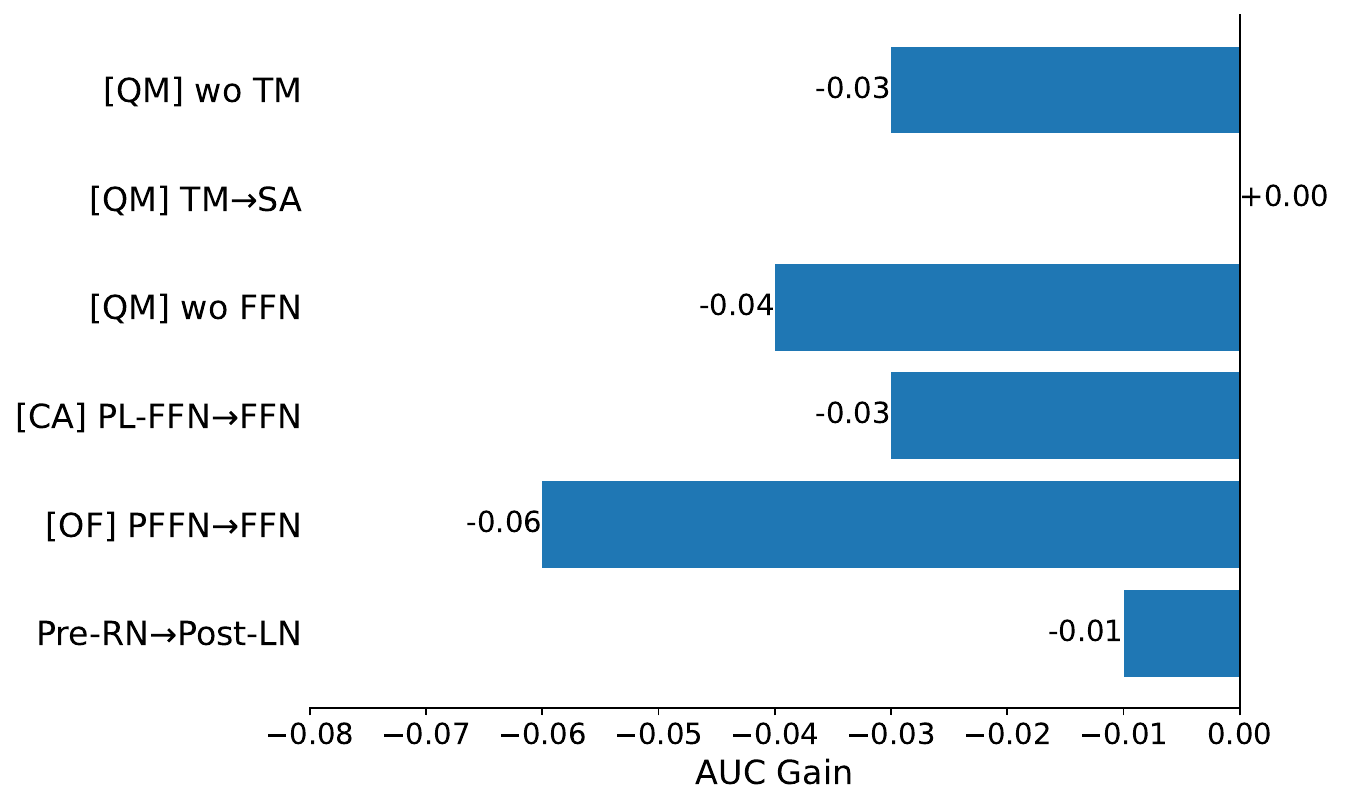}
    \vspace{-0.3cm}
    \caption{The ablation study on modules in MixFormer. The AUC gain is compared with the proposed MixFormer-small. QM, CA, OF, HM, and SA are abbreviations of Query Mixer, Cross Attention, Output Fusion, HeadMixing, and SelfAttention, respectively. PT-FFN and PFFN denote per-layer FFN and per-head FFN, respectively.}
    \label{fig:ablation}
    \Description{The ablation study on modules in MixFormer. The AUC gain is compared with the proposed MixFormer-small. QM, CA, OF, HM, and SA are abbreviations of Query Mixer, Cross Attention, Output Fusion, HeadMixing, and SelfAttention, respectively. PT-FFN and PFFN denote per-layer FFN and per-head FFN, respectively.}
    
\end{figure}

To validate the effectiveness of each module proposed in MixFormer, we conduct a series of experiments by replacing or deleting some designs, which are listed below:
\begin{itemize}[leftmargin=*]
    \item {[QM] wo HM}: deleting HeadMixing in Query Mixer;
    \item {[QM] HM$\to$SA}: replacing HeadMixing with the SelfAttention module in Query Mixer;
    \item {[QM] wo FFN}: deleting per-head FFN in Query Mixer;
    \item {[CA] PL-FFN$\to$FFN}: replacing per-layer FFN of action projection with a shared FFN in CrossAttention.
    \item {[OF] PFFN$\to$FFN}: replacing per-head SwiGLUFFN with a head-shared SwiGLUFFN in Output Fusion;
    \item Pre-RN$\to$ Post-LN: replacing the pre-RMSNorm with the post-LayerNorm in each block;
\end{itemize}
The ablation results are shown in Figure~\ref{fig:ablation}. 
The results demonstrate that both HeadMixing and the per-head FFN in the Query Mixer contribute substantially to performance improvements, underscoring the importance of high-order feature representations when constructing queries for sequence modeling. Notably, replacing HeadMixing with the computationally more expensive self-attention does not yield observable performance gains, which empirically validates the efficiency and effectiveness of zero-cost HeadMixing when handling heterogeneous heads in recommendation.
Besides, the results show that, within cross-attention blocks, the use of per-layer FFNs yields more differentiated representations at various layers, contributing to improved model expressiveness without increasing computation cost.
On the output side, upgrading the standard FFN to a per-head FFN further enhances the model's expressive capacity over heterogeneous heads without increasing FLOPs, leading to a clear performance improvement. Overall, these results indicate that the proposed refinements tailored for recommendation systems achieve favorable improvements for both effectiveness and efficiency.

\subsection{Scaling Analysis}
To assess the co-scaling capability of the proposed model, we evaluate its performance by scaling dense parameters with fixed inputs and by increasing sequence length under a fixed model size.
\subsubsection{Scaling of Dense}
\begin{figure}[t]
    \centering
    \includegraphics[width=1.0\linewidth]{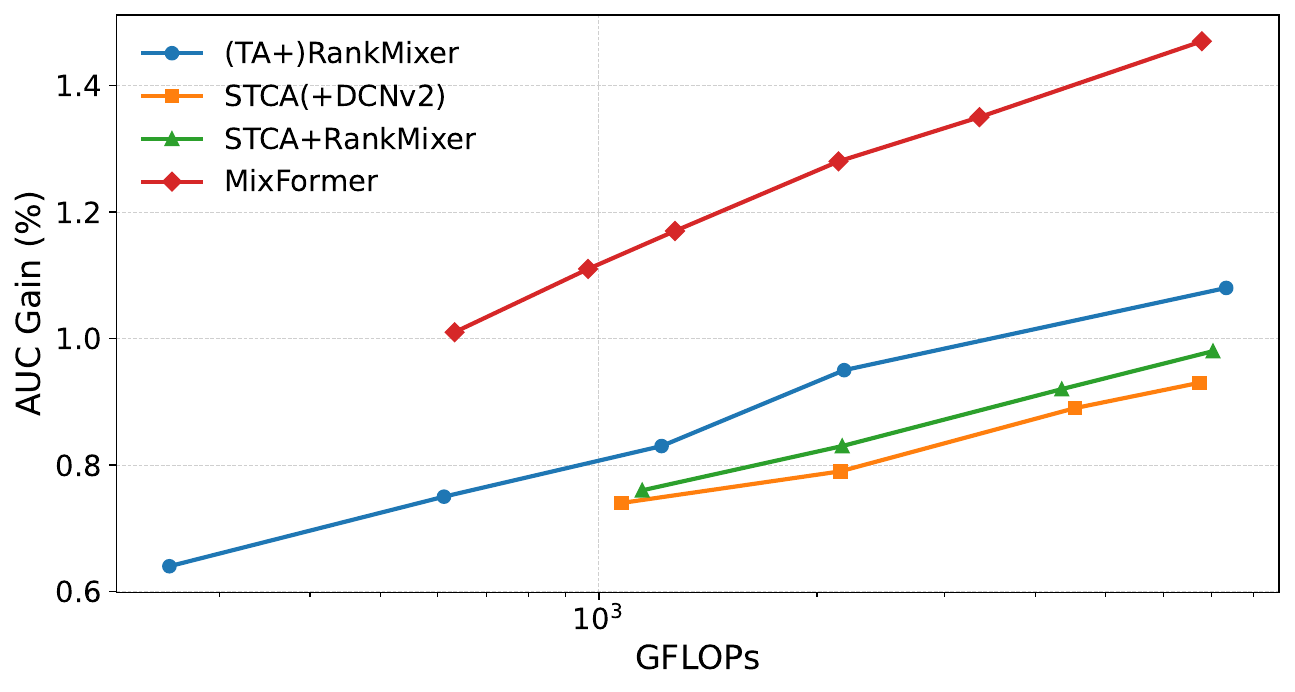}
    \vspace{-0.3cm}
    \caption{Scaling study over FLOPs. The sequence length is fixed at 512. $(A+)B$ denotes the size of module A is fixed while the size of B is scaled.}
    \label{fig:scaling_flops}
    \Description{Scaling study over FLOPs. The sequence length is fixed at 512. $(A+)B$ denotes the size of module A is fixed while the size of B is scaled.}
\end{figure}

Since the computational cost of sequence modeling is highly sensitive to sequence length, comparing dense scaling behavior based solely on parameter count is unfair to non-sequential models. In practice, serving cost is dominated by FLOPs; thus, Figure~\ref{fig:scaling_flops} reports AUC gain as a function of FLOPs.
We compare a state-of-the-art sequential model STCA (with a lightweight DCNv2 head), a non-sequential RankMixer model (with single-layer Target Attention), and their 1:1 FLOPs combination. Under a fixed sequence length setting, scaling RankMixer yields larger marginal AUC gains than scaling the sequential component, highlighting the importance of target-item feature interaction when sequence length is fixed. The combined baseline exhibits a clear trade-off between the two.
In contrast, MixFormer achieves a larger intercept and a competitive scaling slope, benefiting from its unified parameterization and deep interaction between sequential and non-sequential components. As a result, MixFormer consistently outperforms other designs across different FLOPs budgets.

\begin{figure}[t]
    \centering
    \includegraphics[width=1.0\linewidth]{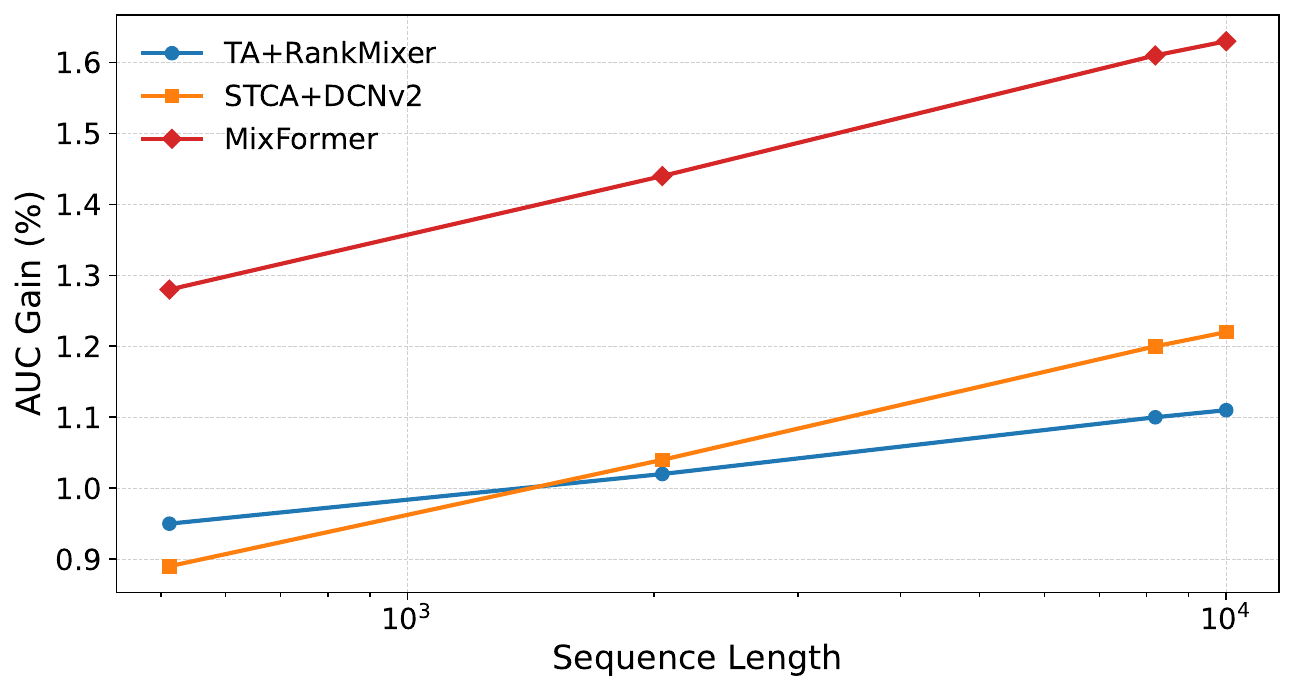}
    \vspace{-0.3cm}
    \caption{Scaling study over sequence length. The sequence length scales among \{512, 2048, 8192, 10000\}.}
    \label{fig:scaling_seq}
    \Description{Scaling study over sequence length. The sequence length scales among \{512, 2048, 8192, 10000\}.}
    
\end{figure}

\subsubsection{Scaling of Sequence}
From another perspective, we examine the sequence-length scaling behavior of different methods under a fixed dense parameter budget. For fairness, we select several model configurations with comparable FLOPs at a sequence length of 512, which are highlighted by circles in Figure~\ref{fig:scaling_flops}. We then scale the input sequence length following~\cite{guan2025make}, evaluating behavior sequences of lengths \{512, 2,048, 8,192, 10,000\}. The results are shown in Figure~\ref{fig:scaling_seq}.
Interestingly, the scaling trend with respect to sequence length exhibits an opposite pattern to dense scaling. Sequential models such as STCA, which allocate more computation to the sequence component, benefit more significantly from longer sequences than non-sequential models, consistent with the observations in~\cite{guan2025make}. Notably, MixFormer achieves a scaling slope comparable to the state-of-the-art STCA, benefiting from its unified parameterization across sequential and non-sequential components.

Overall, we observe that models with independently parameterized sequential and non-sequential components face a pronounced co-scaling trade-off. Under a limited computational budget, such models must carefully allocate capacity between the non-sequential modules with higher FLOPs efficiency and the sequential modules that benefit more from sequence-length scaling. In contrast, owing to its integrated parameter design, MixFormer exhibits state-of-the-art scaling behavior in both dense scaling and sequence scaling, further validating the effectiveness of the architecture.

\begin{figure}[t]
    \centering
    \includegraphics[width=1.0\linewidth]{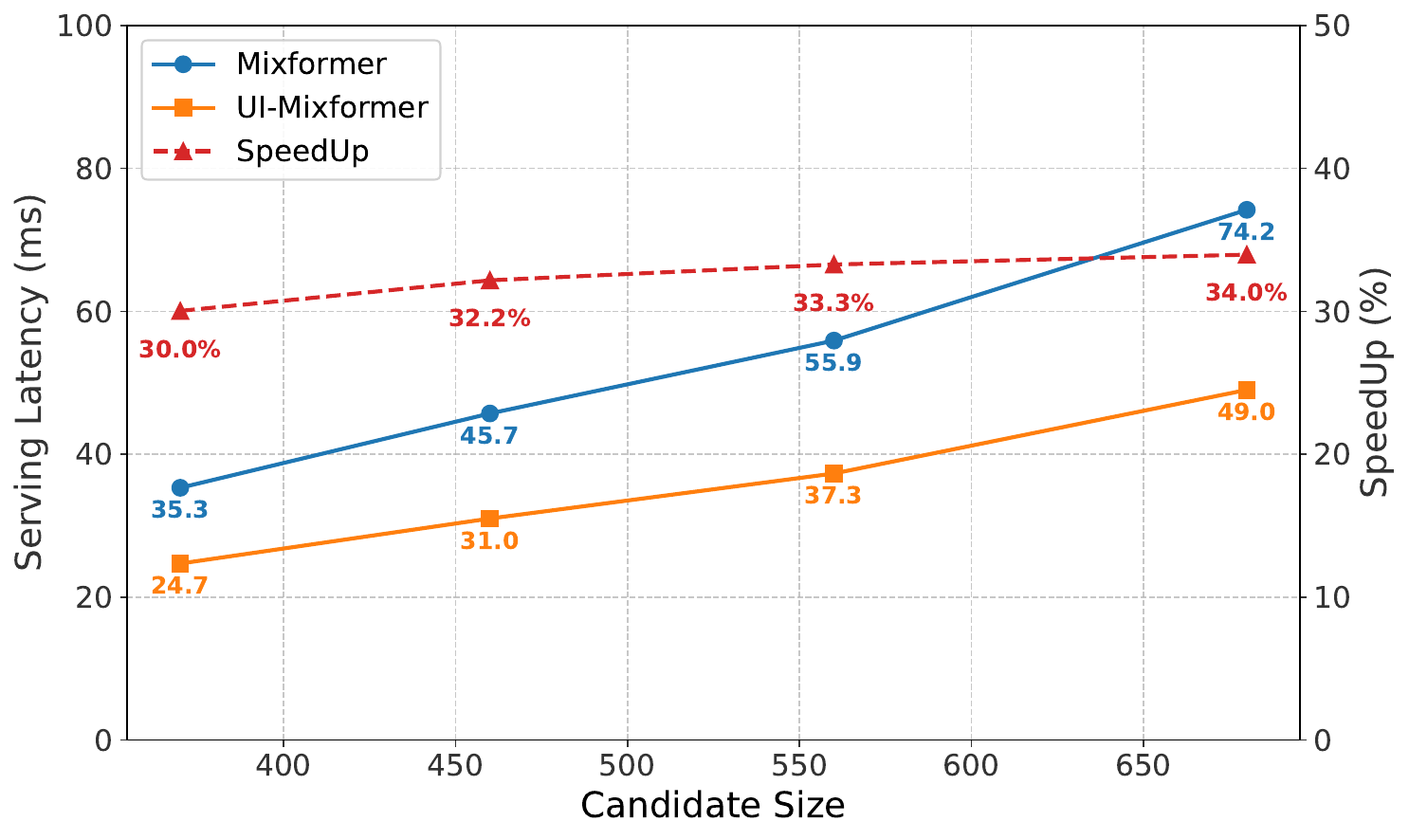}
    \vspace{-0.3cm}
    \caption{Serving Latency and Inference Speedup of Mixformer and User-item decoupled Mixformer (UI-Mixformer) under Different Candidate Sizes. }
    \label{fig:speed_up}
    \Description{Serving Latency and Inference Speedup of Mixformer and User-item decoupled Mixformer (UI-Mixformer) under Different Candidate Sizes. }
    
\end{figure}

\subsection{Serving Latency Anaysis}
We also conduct serving latency tests to validate the efficiency of the user-item decoupling strategy. As shown in Figure \ref{fig:speed_up}, when combined with Request Level Batching, the user-item decoupled MixFormer achieves over 30\% serving speedup.
Moreover, as the number of candidate items to be predicted in the ranking stage scales up, GPU utilization approaches saturation, which further worsens computational bottlenecks. Fortunately, our user-item decoupling design shares user-side computation across multiple candidates within a single request. This results in a far more moderate latency growth for our decoupled MixFormer compared to the original model, improving serving efficiency with large candidate sizes.

\subsection{Online A/B Tests}

To verify the universality of MixFormer as a co-scaling recommendation model framework, we conducted online experiments in the core scenarios of personalised ranking—\emph{feed recommendation} in two Apps.
For feed–Recommendation, we monitor the following key performance indicators:
\emph{Active Days} is the average number of active days per user during the experiment, a substitute for DAU growth; 
\emph{Duration} measures cumulative stay time on the App;
\emph{Finish/Like/Comment:} User's complete plays, likes, and comments.

We conduct online A/B test experiments compared with previous baselines: stacked STCA$\to$RankMixer with more than 1B parameters. The observation of A/B test results on Feed Recommendation for two-weeks are shown in Table~\ref{tab:online_ab}, and the improvements are still increasing indicating the gains have not yet saturated.

\section{Conclusion}
In this paper, we propose a novel Transformer-style unified architecture tailored for recommendation systems. The proposed model integrates the two core components of modern recommender systems—sequence modeling and feature interaction—into a single, unified parameterization, enabling an efficient and easily co-scalable design. Furthermore, we introduce substantial user-item decoupling optimization that explicitly leverage the request-level computation reduction to improve computational efficiency.

Extensive experiments demonstrate that the proposed architecture consistently achieves superior effectiveness and efficiency compared to strong baselines. We further analyze the scaling effect of the model over FLOPs and sequence length, and conduct large-scale online A/B tests in industrial environments, which provide compelling evidence of the model's co-scaling potential and practical effectiveness in real-world recommendation systems.

\newpage


\bibliographystyle{ACM-Reference-Format}
\balance
\bibliography{sections/citations}

\appendix










\end{document}